\documentclass[conference]{IEEEtran}

\usepackage[letterpaper, left=1in, right=1in, bottom=1in, top=0.75in]{geometry}
\usepackage{graphicx}
\usepackage{latexsym}
\usepackage{amsfonts}
\usepackage{amssymb}
\usepackage{amsbsy}
\usepackage{color}
\usepackage{amsmath}
\usepackage{multicol}
\usepackage{float}
\usepackage{subfigure}
\usepackage{algorithm}
\usepackage{algorithmic}
\usepackage[dvips]{epsfig}

\usepackage[nodisplayskipstretch]{setspace}
\usepackage{booktabs}
\usepackage[mediumspace,mediumqspace,Grey,squaren]{SIunits}
\begin{document}

%\IEEEoverridecommandlockouts \IEEEpubid{\makebox[\columnwidth]{ 978-1-5386-3531-5/17/
%\$31.00~\copyright~2017 IEEE \hfill} \hspace{\columnsep}\makebox[\columnwidth]{ }}

\title{Enhancing TCP End-to-End Performance in Millimeter-Wave Communications}

\vskip -10pt

\author{\IEEEauthorblockN{Minho Kim\IEEEauthorrefmark{1}, Seung-Woo Ko\IEEEauthorrefmark{2}, and Seong-Lyun Kim\IEEEauthorrefmark{1}
} 
%\\ Email: swko@eee.hku.hk, huangkb@eee.hku.hk} 
\IEEEauthorblockA{\IEEEauthorrefmark{1}School of EEE, 
Yonsei University, Seoul, Korea}
\IEEEauthorblockA{\IEEEauthorrefmark{2}Dept. of EEE, 
The University of Hong Kong, Hong Kong\\
Email:  \{mhkim, slkim\}@ramo.yonsei.ac.kr, swko@eee.hku.hk}
}

\vskip -10pt

\maketitle

\begin{abstract}
Recently, millimeter-wave (mmWave) communications have received great attention due to the availability of large spectrum resources.
 Nevertheless, their impact on TCP performance has been overlooked, which is observed that  the said TCP performance collapse occurs owing to the significant difference in signal quality between LOS and NLOS links. 
We propose a novel TCP design for mmWave communications, a \emph{mmWave performance enhancing proxy} (mmPEP), 
enabling not only to overcome TCP performance collapse but also exploit the properties of mmWave channels. 
The base station installs the TCP proxy to operate the two functionalities called \emph{Ack management} and \emph{batch retransmission}. 
Specifically, the proxy sends the said early-Ack to the server not to decrease its sending rate even in the NLOS status.  
In addition, when {a packet-loss is detected}, the proxy retransmits not only lost packets but also the certain number of the following packets expected to be lost too. 
It is verified by ns-3 simulation that compared with benchmark, mmPEP enhances the end-to-end rate and packet delivery ratio by maintaining high sending rate 
{with} decreasing the loss recovery time.
\end{abstract}
\vskip -10pt

\section{Introduction}
\vskip -5pt
Due to the availability of large untapped spectrum resources, 
millimeter-wave (mmWave) communications have become the hottest research topic to cope with the dramatic traffic explosion \cite{Rappaport}-\cite{jhpark}. 
Recent studies have shown that the achievable downlink capacity in mmWave networks is in the order of giga bits per second (Gbps) \cite{multiGbps1}.
Nevertheless, limited attention has been paid to Transmission Control Protocol (TCP) performance over mmWave networks
though most end-user services are based on TCP e.g., video streaming, file transfer, and web browsing.
It is of our great interest to uncover a TCP operation over mmWave channels.

\begin{figure}
\centering
\includegraphics[angle=0, width=0.40\textwidth]{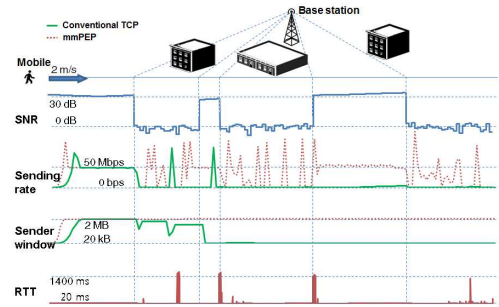}
\caption{Collapse of TCP performance in mmWave communications. The sending rate of a server is limited to 50 Mbps.}
\label{fig:collapse}
\vskip -20pt
\end{figure}

Fig. \ref{fig:collapse} briefly illustrates the TCP performance collapse phenomenon over mmWave channels via our developed end-to-end mmWave ns3-based simulator. 
Given that mmWave signals are sensitive to blockage, 
there is a substantial received signal strength difference between line-of-sight (LOS) and non-line-of-sight (NLOS) links. 
It would be highly problematic because a current TCP protocol does not adapt well 
to such high signal-strength variability. 
Specifically, when a mobile enters an NLOS region, it is not able to receive packets due to low signal strength of NLOS mmWave signals.
This leads to abnormal spikes in round-trip-time (RTT), reducing TCP sending rate to the minimum. 
Due to the current TCP slow-start mechanism, it takes a long time to recover the sending rate to the full even when a mmWave channel resumes LOS status.

Motivated by this observation, this paper proposes a novel TCP design, \emph{mmWave performance enhancing proxy} (mmPEP). The proxy exploits and overcomes the high variability in bandwidth of mmWave channels in LOS-NLOS transitions. 
It is added to a base station capable of temporarily storing packets until a mmWave channel resumes LOS status, whereby the BS then instantly transmits any stored packets.
It comprises the following two operations:

\begin{itemize}
\item \textbf{Ack Management}: The base station sends an Acknowledgement (Ack) to the server in advance before forwarding the corresponding packet to the mobile. 
The Ack sent by the base station is labelled as an \emph{early-Ack} to distinguish it from an Ack normally sent by the mobile to the server. 
The early-Ack operation enables the server to send packets without decreasing its sending rate; that is, the server forwards packets to the base station regardless of wireless channel status.
Note that the base station can send the early-Ack only if available space to store packets exists. 
The base station removes cached packets upon receiving corresponding Acks from the mobile.
In addition, the proxy operates flow-control to avoid an overflow of packets at the base station.

\item \textbf{Batch Retransmission}: 
The proxy contains an operation known as {\it batch retransmission,} whereby, upon detection of packet loss, the base station simultaneously sends not only any lost packets but also a certain number of following packets that are expected to be lost too. Due to the in-order delivery policy of TCP, a receiver must hold all of its existing packets until the arrival of all prior lost packets. This holding period is known as a loss-recovery period -- a period known to result in TCP performance degradation. Though batch retransmission shortens such a period, it is at the expense of unnecessary transmissions of packets already been delivered. Thus, the size of a batch to be retransmitted should be determined carefully according to the properties of mmWave channels.
\vskip -4pt
\vskip -1pt
\end{itemize}

The primary goal of this paper is to provide useful guidelines of TCP design to unleash the full potential of mmWave communications. In recent research, some practical designs of mmWave communications, such as frame structure \cite{mmwave_frame} and MAC layer \cite{mmwave_mac} have been addressed, but
there has been relatively less effort on TCP performance over mmWave networks.
TCP in mmWave communications is addressed in~\cite{TCP_NYU}, but it does not consider  
high lossy channel of NLOS status where a base station cannot transmit packets and the buffer size of Radio Link Control (RLC)  is relatively larger than the practical design value. 
The proposed mmPEP is designed to operate stably in these practical environments. 
In \cite{bufferbloat}, it is claimed the said dynamic receive window approach at the mobile can resolve the TCP performance collapse, 
but it requires the additional channel bandwidth information which can be changed rapidly based on a scheduling policy of a base station. 
On the other hand, the proposed mmPEP does not require any modification of mobiles.  
  
There have been trials at modifying Ack operation and packet retransmission with the aim of enhancing TCP performance \cite{Snoop}--\cite{RFC3135}. 
The local retransmission scheme at the base station is proposed in \cite{Snoop} to reduce the retransmission delay. The authors of \cite{M-TCP} suggest a way to manage Ack operations of wired and wireless links separately to cope with frequent disconnections of wireless links. In \cite{Snoop} and \cite{M-TCP}, however, when mmWave channel is changed from NLOS to LOS, the base station cannot send many packets because the packet arrival rate to the base station is slow due to small sending window at the server. The authors of \cite{RFC3135} propose the said PEP that enables to cache packets in advance by sending an early-Ack. However, the PEP retransmits only one packet when the corresponding duplicated Ack arrives. It is inefficient in mmWave channel because it takes too long time to retransmit many lost packets, bringing about frequent time-out failure of packet delivery. On the other hand, the proposed mmPEP can utilize a large capacity of mmWave channels 
by enabling the transmissions of numerous packets and the fast recovery time of lost packets via Ack management and  batch retransmission, respectively. 

The rest of paper is organized as follows. The system model is presented in Section II. The main functionalities and protocol stack of the mmPEP are presented in Section III.  Simulation results are presented in Section IV followed by concluding remarks in Section V. 
\vskip -5pt
\vskip -5pt

\section{System Model}\label{system}
\vskip -5pt
We consider a downlink network consisting of a server, a mobile, and a base station.
The server and the mobile are at the end of the links as TCP sender and TCP receiver respectively.
The base station is located at the boundary between wired and wireless links.
We assume that the wired link is error-free, to focus more on the impact of wireless channels.

The server and the mobile use conventional TCP.
When the mobile receives a data packet from the server, it forwards an Ack packet to the server as a response to the packet reception.
The Ack packet includes a sequence number that is the last in-order delivered packet to the mobile.

\vskip -5pt

\begin{figure}
\centering
\includegraphics[angle=0, width=0.35\textwidth]{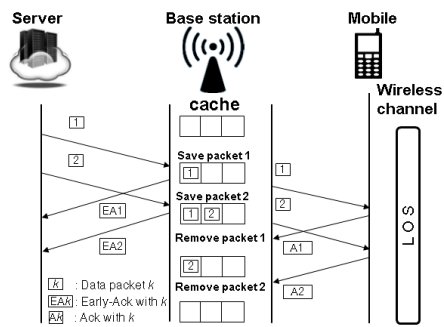}
\caption{An example of the Ack management of mmPEP. Upon receiving packets from the server, the BS stores them in its cache and sends corresponding early-Acks to the server without awaiting the reception of Acks from the mobile. When Acks arrive from the mobile, the base station removes the corresponding stored packets from its cache.} \label{fig:EA_example}
\vskip -5pt
\vskip -5pt
\vskip -5pt
\end{figure}

\section{MmWave Performance Enhancing Proxy}\label{mmPEP}
\vskip -5pt

In this section, we describe the proposed mmPEP in detail. 
Conventionally, when a packet-loss occurs, a server shrinks its sending rate to prevent consecutive packet-losses.
However in mmWave communications, this type of operation may not be necessary, 
because a wireless mmWave channel is likely to regain LOS status within a few seconds.
The proposed mmPEP enables a server to send packets without decreasing its sending rate while guaranteeing successful packet delivery to a mobile through its Ack management and batch retransmission.

\vskip -3pt
\vskip -3pt
\subsection {Ack Management}
\vskip -5pt

Fig. \ref{fig:EA_example} represents a graphical example of the overall Ack management of mmPEP. 
When the base station receives a packet from the server, 
it stores the packet in its cache and sends a corresponding early-Ack to the server. 
Upon verifying successful packet delivery through an Ack from the mobile, 
the base station removes the packet in its cache, and discards the received Ack — since the server already believes the packet to be delivered successfully because of the early-Ack.

We note that the format of an early-Ack is the same as the Ack. The server interprets an early-Ack as a successful packet reception at the mobile, and continues to forward packets to the base station. 
If the status of a mmWave channel is LOS, the base station transmits these packets to the mobile immediately. On the other hand, if the status of a mmWave channel is NLOS, then the base station keeps these packets in its cache until LOS status is resumed. 
In other words, early-Ack operation does not work properly if the cache is full. The base station is thus required to control the server's sending rate by the flow-control like the TCP receiver at the mobile.

Concerning flow control, the base station fills in the Receiver Window (RWIN) field of an early-Ack based on its available cache size.
If its cache is full occupied, then the server stops sending data packets to the base station.
In such a scenario, the base station must now wait until it receives Acks from the mobile, whereby it can then remove the corresponding packets and free up cache space. When the base station has unoccupied cache space, it resends the last known sent early-Ack but with an updated RWIN entry to notify the server that it is now available to receive more packets.

\vskip -3pt
\vskip -5pt

\subsection{Batch Retransmission}
\vskip -5pt
\begin{figure}
\centering
\includegraphics[angle=0, width=0.35\textwidth]{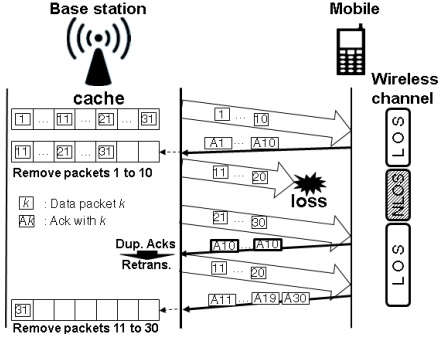}
\caption{An example of the batch retransmission of mmPEP. When packets are lost, the mobile sends duplicated Acks. Then, the base station simultaneously sends not only any lost packets but also a certain number of following packets that are expected to be lost too.} \label{fig:Retransexam}
\vskip -10pt
\vskip -10pt
\end{figure}

{Fig. \ref{fig:Retransexam} illustrates batch retransmission of mmPEP, showing that the base station retransmits many cached packets when it detects a duplicated Ack or a retransmission timer expires\footnote{{RLC layer has its own buffer to retransmit lost packets \cite{RLC}. To avoiding duplicated retransmission, the batch retransmission by the timer expiration is triggered only when the buffer of RLC is empty. We observe the batch retransmission is sometimes triggered by a timer when NLOS lasts more than 15 seconds, in which all packets in flight are lost.}}.}
The use of batch retransmission is inspired by the fact that mmWave channels have a large capacity. However, the vast packet transmission results in the losses of sequential packets when the status of a mmWave channel changed from LOS to NLOS.
To reduce the loss-recovery delay, the base station simultaneously retransmits a certain number of cached packets without receiving their each duplicated Ack. 
The priority of any batch retransmission is higher than that of any initial transmissions.

\begin{figure}
\centering
\includegraphics[angle=0, width=0.40\textwidth]{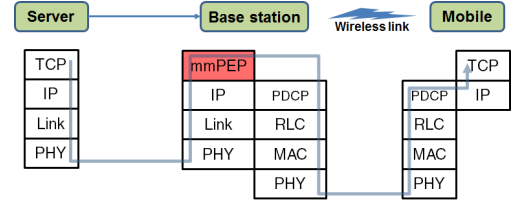}

\caption{Protocol stack design and downlink data flow.} \label{fig:protstack}
\vskip -10pt
\vskip -10pt
\end{figure}

\begin{figure*}
    \centering
    \subfigure[Rate]{
     \includegraphics[width=3.0in]{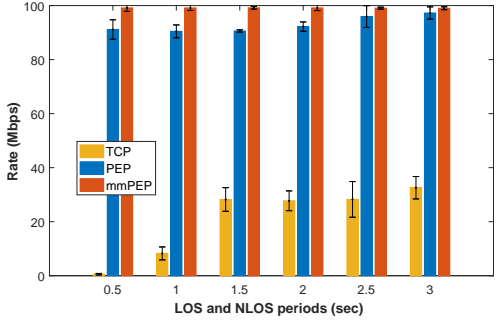}
     \label{Graph1_1}
     }  
    \subfigure[Delivery ratio]{
      \includegraphics[width=3.0in]{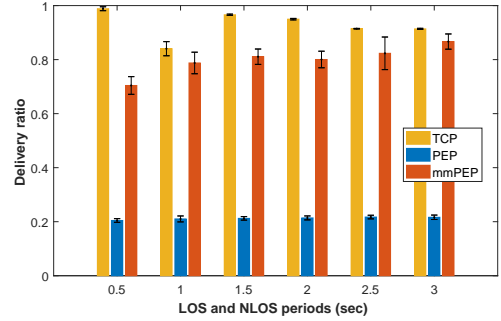}
      \label{Graph1_2}
      }
\vskip -10pt
      \caption{Average rate and delivery ratio with different TCP proxy implementations when LOS and NLOS durations are changed from $0.5$  to $3$ seconds. The ratio between LOS and NLOS durations is fixed to one. The maximum sending rate and RWIN size are  $100$ Mbps and $6$ MB, respectively.  }

    \label{Graph1}
\vskip -20pt
\end{figure*}

Since batch retransmission may incur redundant transmissions, it is important to determine carefully an appropriate number of packets to be resent in each case.
To this end, we refer to a feedback mechanism in the physical layer known as hybrid automatic repeat request (HARQ). 
The reception of a duplicated Ack corresponds to failure of several HARQ retransmissions. 
Specifically, a predetermined time gap exists between packet transmission and arrival of a corresponding HARQ feedback. 
In any given HARQ feedback gap, a sequence of packets is likely to be repeatedly transmitted because NLOS status lasts for a while (at least a few milliseconds) based on the length of the blockage
%\footnote{In our architecture, the reception of a duplicated Ack is regarded as the NLOS status of the mobile.}.
From this insight, once the duplicated Ack is detected in TCP, we regard all corresponding HARQ transmissions as failures for the duration of a HARQ feedback delay.
The total number of batch retransmission is calculated as follows: 
\vskip -15pt
\begin{align}\label{Eq:EstimatedRetransmissionNumber}
\textrm{The number of lost packets}=\alpha \beta(1+\gamma),
\end{align}
where $\alpha$ is the average number of transmitted packets per slot when a channel is available,    
 $\beta$  (in slots) is the timing gap between the occurrence and notification of HARQ packet-loss, and $\gamma$ is the portion that packets are lost beyond  $\alpha\beta$. {Note that these parameters depend on the system model we consider. For example, the requirement of ultra-high-definition (UHD) video is 100 Mbps \cite{4K}, corresponding to $\alpha$=1.12 packets/slot in Long Term Evolution (LTE) where the packet size is approximately 1400 bytes. In addition,  
$\beta$ is 10 slots based on the mmWave physical layer proposed in \cite{mmwave_frame}. Given the  
$\alpha$ and $\beta$, we can derive $\gamma$ based on the assumption that the number of arrival packets during $\beta$ slots follows the normal distribution. It is obvious that the average number of lost packets is $\alpha\beta$ because all arrived packets during the $\beta$ slots are lost. Then $\gamma$ is calculated by the following equation:}
\vskip -18pt
\begin{equation}
\gamma  = \inf \left\{ {\omega \in  \mathbb{R} :\frac{1}{2}\left[ {1 + {\rm{erf}}\left( {\frac{{\alpha \beta \omega}}{{\beta \sigma \sqrt 2 }}} \right)} \right] \ge 1- \epsilon} \right\},\notag
\end{equation} 
{where $\rm{erf}(x)$ is the error function, $\sigma$ is the standard deviation of the arrived packet number, and $1-\epsilon$ represents the error range. Given $\sigma = 0.1$ obtained from the empirical result, the value $\gamma$ approximately becomes $0.2$ when $1-\epsilon$ is $0.99$.}

\vskip -3pt
\vskip -5pt

\subsection{Applying to Long Term Evolution System}
\vskip -5pt

Fig. \ref{fig:protstack} depicts a modification of an LTE protocol stack to apply mmPEP.
It is shown that a separate layer is required for mmPEP — situated between the IP and PDCP layers.
The separate layer is able to monitor TCP and IP packets, both of which are essential to the operations of Ack management and batch retransmission.
It might be argued that the Radio Link Control (RLC) layer is a possible candidate for the location of mmPEP since it contains memory for fast MAC packet retransmissions. \cite{RLC}.   
However, the RLC layer is unable to access a TCP packet header.

\begin{figure*}[t!]
    \centering
    \subfigure[Rate]{
     \includegraphics[width=3.0in]{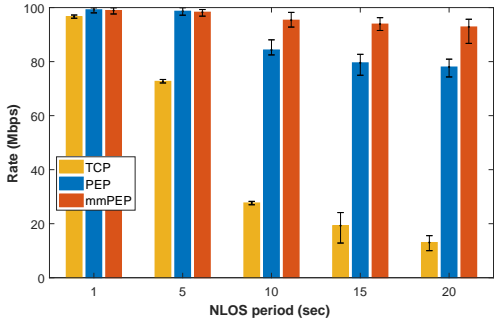}
     \label{Graph2_1}
     }  
    \subfigure[Delivery ratio]{
      \includegraphics[width=3.0in]{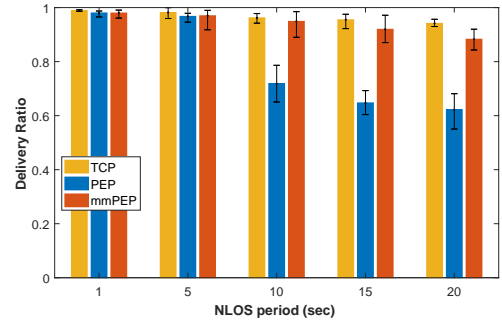}
      \label{Graph2_2}
      }
\vskip -10pt
      \caption{Average rate and delivery ratio with different TCP proxy implementations under long LOS regime where NLOS is changed from $1$ to $20$ seconds but LOS duration is fixed to $100$ seconds.   The maximum sending rate and RWIN size are  $100$ Mbps and $6$ MB, respectively.  }

    \label{Graph2}
\vskip -20pt
\end{figure*}

\vskip -7pt
\vskip -10pt
\section{Simulation Results}\label{SimulationResult}
\vskip -3pt
This section presents simulation results for evaluating the performance of mmPEP  
in comparison with conventional TCP and PEP \cite{RFC3135}
where Ack management is applied but only packets receiving the corresponding duplicated Acks are 
retransmitted. 
The simulation is based on LENA \cite{lena} that implements LTE using the Network Simulation version-3 (ns-3) \cite{ns-3}. 
For a mmWave radio propagation model in PHY layer and related operations in MAC layer, 
we utilize the mmWave framework suggested in \cite{NYU}
The detailed parameters are described in Table \ref{tab:table1}.

\begin{table}
  \centering
  \caption{Parameter configuration in ns-3 \cite{NYU}.}
\vskip -10pt
  \label{tab:table1}
  \begin{tabular}{ccc}
    \toprule
    Parameter & Value \\
    \midrule
    Carrier frequency & 28 GHz  \\
    System bandwidth & 1GHz\\
    Transmission power & 30 dBm \\
    Number of antennas (Tx/Rx) & 64, 16 \\
    Noise figure & 5 dB \\
    LTE duplexing mode & TDD \\
    TDD pattern & ``CCDDDUUU'' \\
    Slot length &  \unit{125}{\micro\second} \\
    Packet rate ($\alpha$) & 1.12 packets/slot  \\
    HARQ feedback delay ($\beta$) & 10 slots \\
    Additional packet lost portion ($\gamma$) & 0.2 \\
    \bottomrule
  \end{tabular}
\vskip -10pt
\vskip -10pt
\end{table}

We use the two following performance metrics:
\begin{itemize}
\item {\it Rate}: The number of in-order bits that the application layer receives for a second. 
\item {\it Delivery ratio:}  The ratio of packets delivered to the application layer to packets arrived at TCP layer.
\end{itemize}
Note that the above performance metrics are measured at the application layer 
because it is the closest to user interfaces and can reflect Quality-of-Experience (QoE).

We set up a single TCP-NewReno pair of a server and a mobile and the mobile linearly moves with a constant speed and experiences in LOS and NLOS alternately 
depending on how frequently it encounters blockages. %The ratio between LOS and NLOS durations is fixed to one.
Every figure is plotted against different length of LOS and NLOS periods 
that is determined by the blockage environment.

Fig. \ref{Graph1} represents the performances of different TCP implementations when LOS and NLOS durations are changed.
The ratio between LOS and NLOS durations is fixed to one.
As shown in Fig. \ref{Graph1_1}, the proposed mmPEP has the highest rate. 
Especially, the average rate of mmPEP is almost close to the maximum sending rate of the server 
that is fixed to $100$ Mbps in the simulation. 
On the other hand, PEP does not achieve the maximum sending rate 
and the difference becomes larger as LOS period shortens. It means that 
the batch retransmission of mmPEP  can exploit vast bandwidth of mmWave channel efficiently, yielding the near-to-optimal rate. 

Delivery ratios are plotted in Fig. \ref{Graph1_2}, showing that  mmPEP can achieve three times higher delivery ratio than PEP because of different retransmission timer usages. 
%The retransmission timer is determined by the recent RTT measurement. 
In PEP, when NLOS becomes longer, a redundant retransmission by timer can occur because RLC is able to retransmit the corresponding packet if it is in the RLC buffer. In mmPEP, on the other hand, the batch retransmission is triggered by timer expiration only when RLC buffer is empty. The condition plays a role to achieve efficient packet delivery by preventing the redundant retransmissions. It is observed that the conventional TCP achieves the highest delivery ratio because a retransmission is triggered by not a base station but a server, resulting in less trials of packet retransmission than PEP and mmPEP. Nevertheless, the conventional TCP is not an efficient scheme in mmWave communication because its rate is quite low shown in Fig. \ref{Graph1_1}.

%PEP redundant retransmissions by the retransmission timer expiration even though packets are not lost. 
%The number of PEP's retransmissions increases even when small RTT spike of NLOS arises 
%because its retransmission timer is based on RTT between the base station and the mobile.
%On the other hand, the batch retransmission of mmPEP enhances efficiency of packet delivery by reducing unnecessary retransmissions due to RTT spikes of mmWave. 
%Although this unnecessary retransmissions occur in TCP but the sending rate of the TCP shrinks to the minimum so that the amount of the retransmission is low.

Fig. \ref{Graph2} shows the rate and delivery ratio of different TCP implementations under long LOS regime where LOS period is relatively larger than NLOS. Similar tendencies in Fig. \ref{Graph1} are observed but one distinct point is the significant rate gap between PEP and mmPEP when NLOS period becomes larger than 10 seconds. Under NLOS status, the mobile receives out-of-order packets that cannot be delivered to the application layer until the reception of in-order ones. As NLOS lasts more than 10 seconds, some packets should be discarded in PEP because there is no available cache space in the mobile to save the next arrived packets. This delays loss-recovery and resulting in more retransmissions by a timer. On the other hand, the batch retransmission of mmPEP allows to deliver packets without the corresponding duplicated ACKs, making the sequence of in-order packet faster. This difference leads to the gap of delivery ratios when NLOS becomes longer than 10 seconds.

\vskip -10pt

\section{Conclusion}\label{Conclusion}
\vskip -5pt
In this paper, we tackled the issue of TCP performance collapse in mmWave communications where a TCP sending rate falls off and does not recover for a long time.
To this end, we proposed a novel TCP design called mmPEP that operates Ack management and batch retransmission depending on the status of the mmWave channel. 
Specifically, the base station sends an Ack to the server before forwarding the corresponding packet to the mobile, 
and retransmits not only lost packets but also the following ones at once when a channel is turned into LOS. 
Comprehensive simulation based on ns-3 shows that mmPEP achieves not only high data rate but also reliability over mmWave channels.
Although more memory is required proportionally to the number of users in a cell, the mmPEP will be useful especially in ultra-dense networks \cite{jhpark}
where only a few users exists in each cell due to its small coverage.

\section*{Acknowledgement}
\vskip -5pt
This work was supported by 'The Cross-Ministry Giga KOREA Project' grant funded by the Korea government (MSIT) (No.GK17S0400, Research and Development of Open 5G Reference Model)
\vskip -5pt

\ifCLASSOPTIONcaptionsoff
  \newpage
\fi

% that's all folks
\end{document}